# Ultrafast Electron Diffraction with MeV Electron Source from a Laser Wakefield Accelerator

Yu Fang[1], Fei Li[1*], Jianfei Hua[1*], Bo Guo[2], Linyi Zhou[1], Bing Zhou[1], Zhihao Chen[1], Jianyi Liu[1], Zheng Zhou[3], Yipeng Wu[4], Yingchao Du[1], Renkai Li[1], Wei Lu[1,2*]

**MeV ultrafast electron diffraction (UED) is a widely used technology for ultrafast structural dynamic studies of matters in numerous areas. The development of laser wakefield accelerator (LWFA) envisions great potential of advanced all-optical electron source based on LWFA in UED applications. We experimentally demonstrated that an LWFA-based device with a miniaturized permanent magnet beamline can generate and manipulate electron beams suitable for UED. In the beam transmission, the LWFA electron beams with intrinsic short duration stretch due to energy spread and then are compressed by a following double bend achromat. The optimized double bend achromat can make the beamline isochronous such that the arrival time jitter induced by the shot-to-shot energy fluctuation can be eliminated, and allow the advantage of the natural laser-beam synchronization for LWFAs to emerge. With the energy filtering, the beam energy spread can be reduced to 3% (FWHM) while a sufficient amount of charge (11.9 fC) per bunch for diffraction is retained. Start-to-end simulations showed that the bunch length reaches ~30 fs (rms) with the same experimental configuration. Clear single-shot and multi-shot diffraction patterns of single-crystalline gold samples are obtained and the derived lattice constant agrees excellently with the real value. Our proof-of-principle experiments open the door to the detection of ultrafast structural dynamics using MeV LWFA beams, and pave the way for the UED applications with sub-10-fs temporal resolution.**

Achieving direct observation of ultrafast structural dynamics of matter at the atomic scale is of great importance to the fields of solid-state physics, femtosecond chemistry and bioimaging[1-6]. Ultrafast electron diffraction (UED) is a very powerful tool for visualizing atomic and molecular dynamics[7-8] with high spatial (~100pm) and temporal (<100fs) resolution. MeV UED technique has been proposed and actively developed for years. Compared to the electron diffraction using 10s~100s keV electron beams[9-13], the MeV-UED not only greatly improves the temporal resolution by mitigating the broadening of beam length induced by space charge force, but also is more suitable for the detection of thick samples due to the longer mean free path.

The MeV-UED experiments to date are almost based on the traditional radio frequency (rf) accelerators. However, the current rf technique still suffers from two major factors that limit the temporal resolution. One of them is the intrinsic rf jitter which increases the arrival time jitter between the pump laser and the probe beam. The other is the relatively long bunch length (typically tens of femtoseconds to a few picoseconds) of the electron beam caused by the space charge effect in the rf gun. In the past few years, various techniques have been proposed to improve the temporal resolution of MeV-UED such as the rf bunch compression[14-15], the time-stamping technique[15-17], the laser-driven THz bunch compression[18-19] and the double bend achromatic structure[20-21]. Despite the temporal resolution has been significantly improved and less than 50 fs has been achieved[20], the aforementioned approaches still face grand hurdles in continuing to break the 10-fs resolution due to the inherent restriction of rf jitter. THz-based time-stamping method can achieve time synchronization close to 1 fs, but only for single-shot or low repetition rate modes.

Laser wakefield accelerators (LWFAs) have received wide attention in the past decades due to the extremely high acceleration gradient compared against conventional accelerators[22-25]. The unique advantages of LWFAs for UED applications – the ultrashort bunch length and the jitter-free time synchronization – have also been highly pursed. Due to the tiny accelerating structure of tens of $\mu$m, the electron bunches naturally have a fs-level bunch length. The jitter-free synchronization stems from the fact that the plasma wake is directly driven by the laser. These two advantages make LWFAs promising to achieve sub-10-fs resolved UED devices. Moreover, the UED devices based on LWFAs, i.e., the all-optical UED, are much more compact and cost-effective than those based on traditional accelerator techniques.

[1]Tsinghua University, Beijing, China. [2]Beijing Academy of Quantum Information Science, Beijing, China. [3]China Academy of Engineering Physics, Mianyang, Sichuan, China. [4]University of California, Los Angeles, Los Angeles, CA, US.
*Email: lifei2023@tsinghua.edu.cn; jfhua@mail.tsinghua.edu.cn; weilu@mail.tsinghua.edu.cn;

The UED applications have stringent requirement for the energy spread of the electron beams. First, the lower energy spread corresponds to a smaller wavelength dispersion of de Broglie wave and thus a sharper diffraction pattern. Second, the energy spread $\delta E/E$ will lead to the elongation of bunch length $\Delta l$ during the beam transport and hence deteriorate the temporal resolution. The amount of beam length elongation can be evaluated with $\Delta l = R_{56}\delta E/E$ where $R_{56}$ is the matrix element of the transmission matrix from the LWFA exit to the sample.

In recent years, several efforts have been devoted to explore the possibility of using LWFA for UED experiments, including keV-LPA-UED experiments[26-28] and MeV-UED beamline design[29]. In this Article, we demonstrated the proof-of-principle all-optical MeV-UED experiment based on LWFA. The energy spread of the LWFA output electron beams can reach 3% (FWHM) after the beam post-processing with the central energy of 4.27MeV. The LWFA was optimized to be operated at the high beam charge mode such that the peak charge-per-MeV reaches 20 pC/MeV, which is critical to the success of the experiment. Simulation indicates that the bunch length is compressed to 30fs (rms) at the position of the sample after the 0.5-m-long beam transport. We succeed to observe clear diffraction patterns of single-crystalline Au samples. This result lays an important foundation for the application of LWFA to the ultrafast structural dynamics research.

**System layout**

The UED system mainly consists of an LWFA and a dedicated beamline as shown in Fig. 1a. The 800 nm laser pulse with a duration of 24 fs (FWHM) and an energy of 130 mJ is provided by a compact TW Ti:Sapphire laser system. The laser is focused by an off-axis parabolic mirror to a spot size of ~6.5 μm (FWHM) in the front of a supersonic gas jet of nitrogen. The pointing stability of the drive laser is optimized to 2 μrad to generate stable output electron beams. The gas jet of nitrogen is produced by a gas nozzle with a diameter of 1 mm and provides uniform plasma with a density of $4\times10^{19}$ cm$^{-3}$ after the ionization by the laser front.

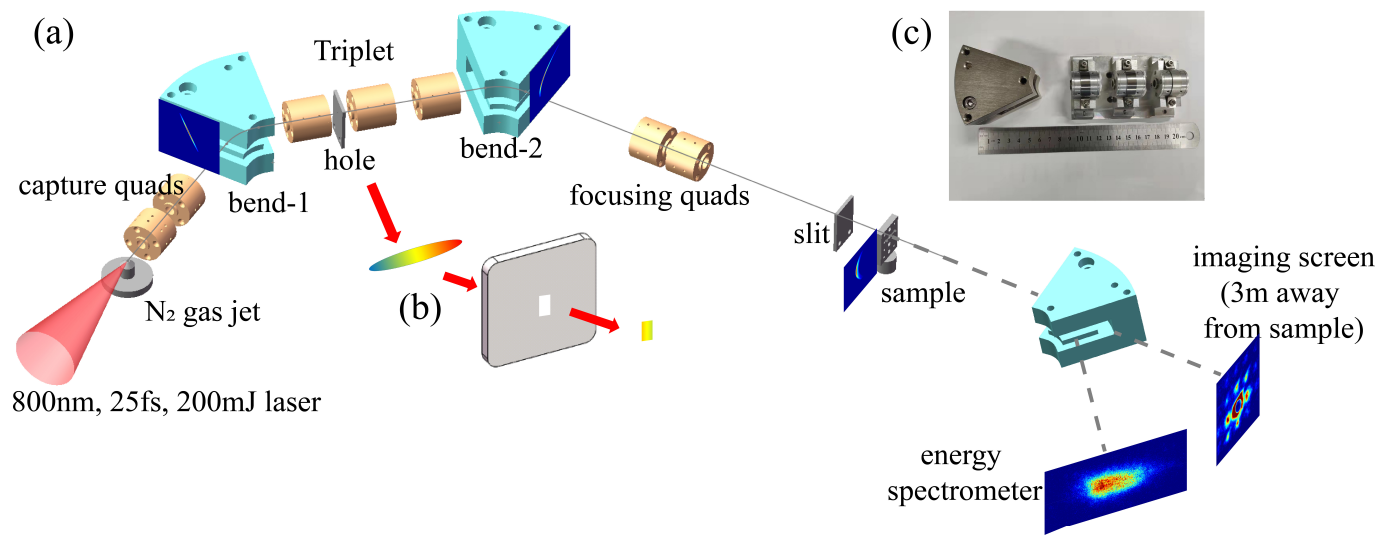

Fig.1. (a) Layout of the all-optical MeV-UED device. The beamline primarily consists of a DBA structure, a permanent quadrupole triplet and (b) a beam energy filter. (c) The photo of the magnets.

When the laser pulse is incident into the plasma, the ponderomotive force excites a large-amplitude plasma wake traveling at close to the speed of light. The inner-shell electrons of nitrogen atoms are trapped by the plasma wake through the ionization injection mechanism, and then accelerated to the desired energy of 4.27 MeV.

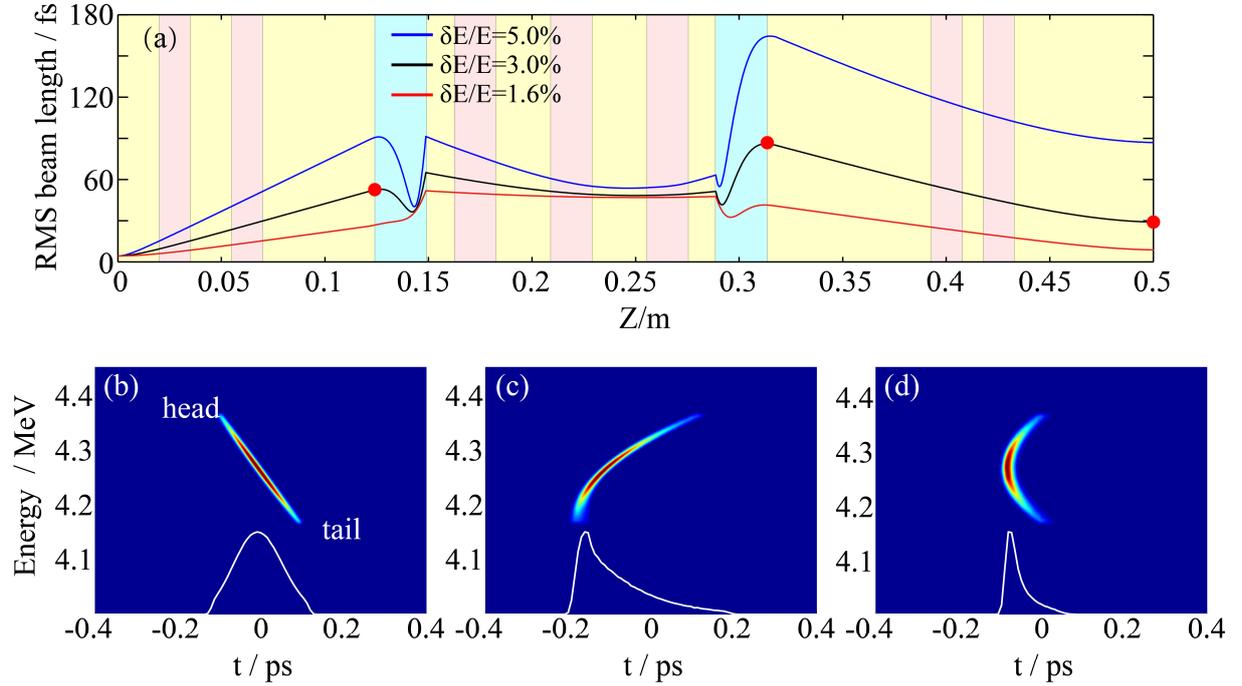

Fig.2 (a) Simulated pulse duration evolution along the beam transport and longitudinal phase spaces of beam at (b) entrance, (c) exit of DBA and (d) sample (the three red dots in Fig.2a) with initial energy spread of 3% and divergence of 2 mrad. Blue (pink and yellow) areas in (a) denote the position of dipoles (quadrupoles and drifts).

The output electron beams initially with a large energy spread and divergence are captured and focused by a doublet of permanent magnet quadrupoles to the double bend achromatic (DBA) structure. The DBA structure consists of two 45° bending magnets and a quadrupole triplet in between, and plays a critical role in the bunch compression and the suppression of the arrival time jitter caused by the shot-to-shot beam energy fluctuation. More specifically, the longitudinal dispersion $R_{56}$ of the first beam coupling section from the LWFA to the entrance of DBA and the second one from the exit of DBA to the sample location are positive. Fig. 2a shows the simulated beam length evolution of an electron beam with central energy of 4.27 MeV, initial divergence of 2 mrad and different energy spread during the beam transport from gas jet to sample. The electron bunch is stretched in the first coupling section and present a positive-chirped longitudinal phase space distribution (the beam head has a higher energy than the beam tail) at the DBA entrance as shown in Fig. 2b. On the contrary, in the DBA structure a particle with higher (lower) energy passes through a longer (shorter) path, leading to a negative longitudinal dispersion $R_{56}$, which enables the bunch compression for the initially positive-chirped beams. We slightly over-compressed the beam in the DBA structure so that the beam has a negative chirp at the DBA exit as shown in Fig. 2c. Then the negatively chirped electron beam passes through the second coupling section and is further compressed via velocity compression until the sample position where the bunch tail catches up with the bunch head, resulting in an uncorrelated, fully compressed longitudinal phase space as shown in Fig. 2d. A well-designed DBA structure can make the whole system (from the LWFA to the sample location) isochronous, i.e., the $R_{56}$ factor of the whole beamline exactly vanishes. In this situation, the bunch length at the sample location is minimized and the arrival time jitter induced by the shot-to-shot energy fluctuations should be fully eliminated.

A 2-mm-thick tungsten(W) rectangular hole with a size of 400×800 μm is deployed inside the DBA structure as a bandpass energy filter, as shown in Fig. 1b. It selects only a small fraction of electrons around the central energy to reduce the beam energy spread down to ~3%, satisfying the requirement of a clear diffraction pattern. The electron beam can also be collimated by this energy filter hole and a beam collimator slit after the second bending magnet. Of particular note is that the single-shot diffraction imaging is a must for some sample-irreversible experiments. In this scenario, appropriate trade-off should be made for choosing the size of slits to obtain sufficient charge per bunch (typically above 1 fC) and thus ensure a good signal-to-noise ratio (SNR). In addition, the energy filter hole and the beam collimator can also crop out the electrons at the bunch head and tail that possess relatively large nonlinear curvature, which will be beneficial for better bunch compression. The imaging system (see the Methods for the detailed set-up) is located 3 m away from the sample location, which is enough to resolve the spatial.

It is noteworthy that we used permanent magnets for all the beamline components to downsize the whole UED device. As a result, the flight path of electron beams from the gas nozzle to the sample is only 0.5 m.

Due to the phase-jitter-free nature of LWFAs and the isochronism of the beamline, the pump-probe synchronization can be considered nearly perfect and the probe beam length becomes the main factor in determining the temporal resolution. Through simulation, the beam length of a MeV electron beam can be compressed to 30 fs (rms) with 3% energy spread selected, and 10fs (rms) beam length with 1.6% energy spread as shown in Fig. 2a, indicating the potential of achieving sub-10 fs temporal resolution. Since the low energy spread is realized by the energy selection of the beamline, the charge-per-MeV becomes the parameter that needs to be optimized specially in order to guarantee sufficient charge for the diffraction (typically >1 fC). For this purpose, we operated the LWFA in the large-charge mode with a relatively large energy spread.

It is worth mentioning that this beamline not only satisfies the central energy of 4.27MeV, but also is suitable for electron beams with similar energy by the position adjustment of the dipoles and quadrupoles.

## Characterization of the electron probes

The profiles of the LWFA beams are monitored at the position of 310 mm away from gas cell. As shown in Fig. 3a, the measured beam size (FWHM) is 17.8mm (horizontal) and 13.42mm (vertical), corresponding to the divergences of 57.4 mrad and 43.3mrad in these two directions.

The quadrupole doublet following the LWFA is used to stabilize the pointing jitter of the electron beams and capture sufficient amount of charge for subsequent energy filtering. After the focusing of the quadrupole magnets, the fluctuation of the transverse beam position at the energy filter hole and collimator slit can be significantly reduced, and thus the beam charge throughput can be improved. As seen from Fig. 3b, the spot size of the electron beams at the imaging screen at the end of beamline is measured to be 1.4 mm (FWHM). As a result of the high throughput, the final beam charge per unit energy is orders of magnitude higher than that of the LWFA experiments driven by a few-cycle laser pulse[25]. This advantage makes sufficient beam charge can remain after the beam truncation, thus making single-shot diffraction imaging feasible even for LWFA electron beams with large energy spread.

As mentioned previously, the LWFA is operated in the high-charge mode to optimize the charge-per-MeV parameter. This mode is realized using the continuous ionization injection which has a relatively large energy spread as shown in Fig. 4a and 4b. The total beam charge before the energy filtering is approximately 45 pC and the energy spread (FWHM) is 70%. After the energy filter and collimator, a small fraction of electrons near the peak energy 4.27 MeV is selected with a measurement error of 1.12% (see the Methods for the details). The typical energy spectrum is shown in Fig. 4c and 4d. The details of the error analysis can be found in the Methods. By counting the gray value of the beam profile images[30], the average bunch charge after the energy filtering is estimated to be 11.9 fC with the shot-to-shot fluctuation of 23% (rms). The measured energy spectrum (FWHM) is ~3%, as shown in Fig. 4d.

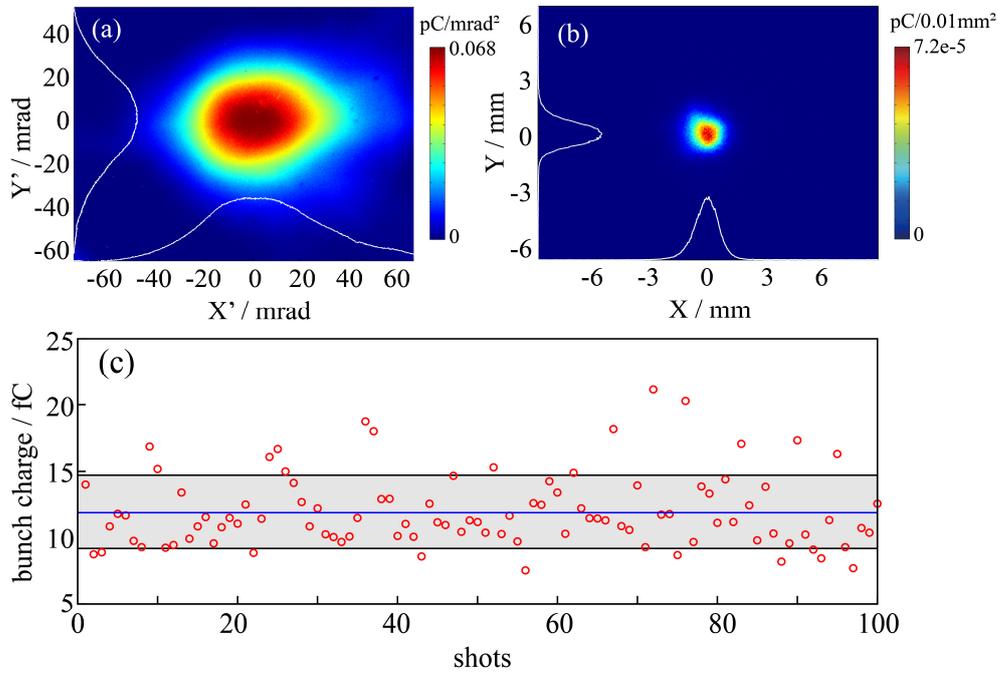

Fig.3. Electron profiles measured (a) 310 mm away from the gas jet without the first pair of quadruples and (b) at the imaging screen located in the end of all-optical MeV-UED beamline. (c) is the bunch charge of continuous 100shots, and the blue line and the gray area indicates the mean value and the RMS fluctuation.

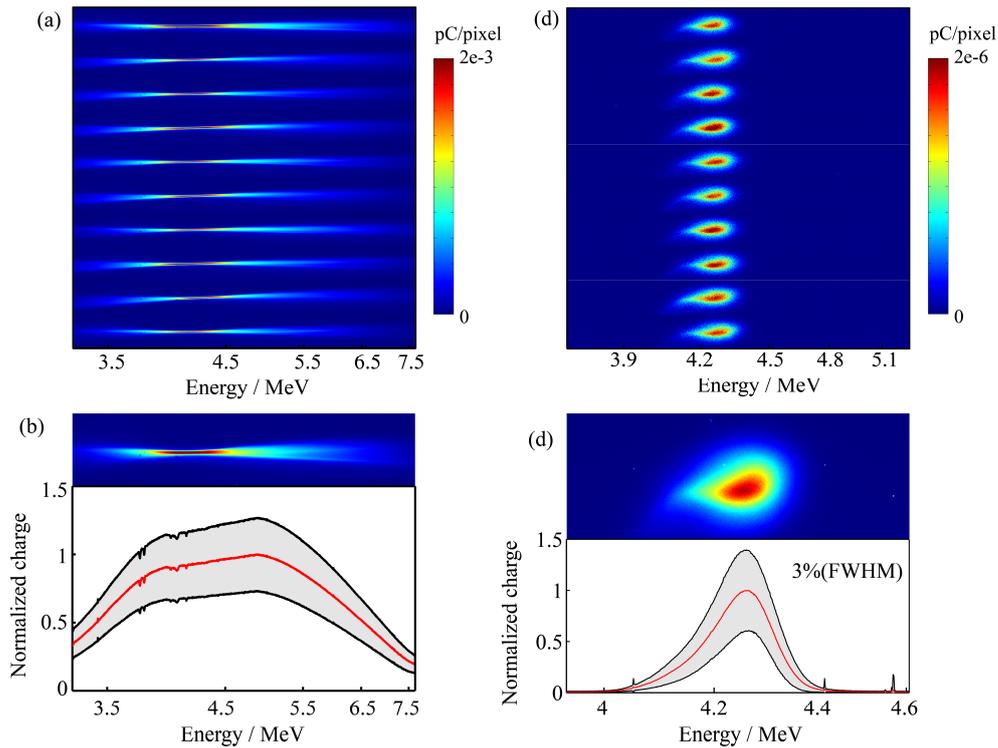

Fig.4 The energy distribution measured of (a-b) the electron source and (c-d) the selected beam bunch. The gray area indicates the RMS fluctuation.

## Proof-of-principle demonstration of ultrafast electron diffraction

We have carried out proof-of-principle diffraction imaging experiments using single-crystalline gold samples and successfully observed clear diffraction patterns in both single-shot and multi-shot operation modes, validating the feasibility of this system for the UED applications. The gold crystal specimen consists of an ultra-thin layer of gold grown epitaxially into single crystals that are oriented to display the lattice spacings of 0.204nm (200 plane) and 0.143nm (220 plane).

Fig. 5a shows the diffraction pattern of the single-shot experiment. The diffraction spots can be well recognized from the background albeit there is still room to further improve the SNR by suppressing the clock induced charge noise of the EMCCD camera. Owing to the focusing of the two sets of quadrupoles before and after the DBA structure, the transverse position fluctuations of the diffraction pattern on the imaging screen are only 0.27 mm and 0.07 mm in the horizontal and vertical directions respectively. The fluctuation amplitude is approximately 1/5 of the diffraction spot. The excellent transverse position stability enables the multi-shot experiment and the high SNR diffraction accumulated by 100 shots can be seen in Fig. 5b.

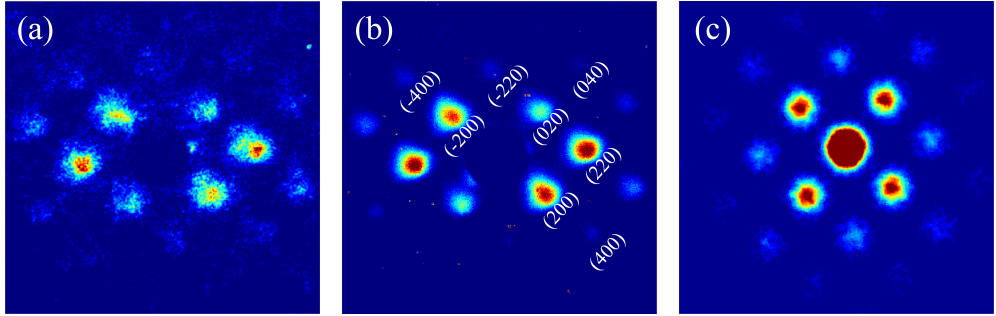

Fig.5 (a) Diffraction images of single-shot, (b) multi-shot (100 shots) and (c) simulation results of single-crystalline gold.

By calibrating the spatial resolution of the imaging screen, the spacing between adjacent diffraction spots is measured to be 3.79 mm ($\pm 10$ μm). With the sample-to-screen distance being 3 m in mind, this corresponds to the diffraction angle $\theta$ of about 1.26 mrad. According to the Bragg diffraction formula $d\sin\theta = \lambda$, the lattice constant $d$ of single-crystalline gold is 0.207 nm with the de Broglie wavelength $\lambda = 0.261$ pm which corresponds to the 4.27 MeV ($\pm 30$ keV) electrons. The final measurement error of the lattice constant is estimated to be 1.5%, which has taken into account the beam peak energy error, the beam energy spread and the optical imaging system error (see the Methods for the details).

**Conclusion**

In conclusion, we have built a compact all-optical MeV-UED device driven by a LWFA and successfully carried out proof-of-principle UED experiments. The energy spread of the electron beams generated by the LWFA can be reduced down to 3% after the energy filter slit while retaining sufficient charge (11.9fC) that is available for single-shot diffraction imaging. By using the DBA structure, the isochronous beamline can compress the beam at the sample position to the greatest extent and theoretically eliminates time jitter induced by the shot-to-shot beam energy fluctuation. The excellent stability of bunch transverse position allows operating the MeV-UED device in the multi-shot mode, and high SNR diffraction patterns are obtained. The clear diffraction pattern of single-crystalline gold sample leads to an accurate measurement of the lattice constant with the spatial resolution of 1.5%. Much better spatial resolution can be anticipated by optimizing the beam quality and stability of the LWFA and suppressing the noise level of the imaging system. Future work also includes carrying out pump-probe detection of ultrafast structure dynamics of various materials. Our experimental demonstration has opened the door to the use of LWFA to probe ultrafast processes in the microscopic world.

## Methods
### Laser wakefield accelerator.

The 800nm laser pulse driving the LWFA was generated from a compact laser system, and then focused to 6.5 μm (FWHM) spot size above the nozzle by an off-axis parabolic mirror with focal length of 300 mm. The energy fluctuation of the laser was ~1% (RMS) and the transverse position fluctuation of the focal spot was ~0.6 μm.

The plasma was generated by photo-ionizing the nitrogen supersonic gas jet ejected from a 0.8 mm diameter nozzle with throat diameter of 0.4mm.

The particle injection process of LWFA can be controlled by tuning the laser intensity, pulse duration and plasma density to obtain stable MeV electron beams. After optimization of the electronic source, the laser pulse duration was compressed to 24 fs (FWHM), and the on-target laser pulse energy was ~128 mJ. A Nomarski interferometer[31], employing a Wollaston prism to divide the beam, is set for online real-time measurement of the plasma density. A relatively optimized backing pressure of the gas jet that leads to stable output was found ~0.1 MPa, which corresponds to a peak plasma density of $4 \times 10^{19}$ cm$^{-3}$.

### Beamline design and beam dynamics simulation.

The UED beamline lattice was firstly designed based on the transfer matrix theory. All transmission elements were designed and manufactured based on the permanent magnet technique. Due to the fringe field effect of magnets, the trajectory of electron beams through these transmission elements, especially the sector bend magnet, deviates somewhat from the prediction of the ideal magnet model. Thus, we mitigated and characterized the fringe fields through finite-element simulations using CST, and have taken the effect into account in the beamline design.

Table1 Transmission element parameters

|  | Bend magnet | Capture quads | Quad. triplet | Focusing quads |
|---|---|---|---|---|
| Strength | 0.434 T | 63.9 T/m<br>27.1 T/m | 34.0 T/m | 34.7 T/m<br>27.5 T/m |
| Length | 28.7 mm | 15 mm | 20 mm | 15 mm |

With particle tracing simulations using GPT, the beamline lattice was further finetuned considering the magnet fringe fields and the space charge effect of electron beams. The test electron beams consist of $10^7$ Gaussian distributed macroparticles with 4.27 MeV central energy, 70% energy spread and 45 mrad divergence angle, verifying the feasibility of the beamline. The space charge effect was solved using Spacecharge3Dmesh algorithm of GPT. We determined the beamline layout and the position of slit and hole by optimizing the duration, energy spread and charge of the electron bunches after the selection by the energy filter and beam collimator. The parameters of transmission elements are shown in Table1.

### Electron beam charge measurement.

The electron beam profiles in the experiment were obtained by an X-ray scintillator screen (PI-200) and an Electron-Multiplying CCD. The bunch charge is calculated by counting the gray value of the captured images. The gray value counts $H$ are determined by the number of electrons $N_e$, the optical transmission efficiency of the imaging system $f_{optical}$, the conversional efficiency of DRZ-high screen $f_{DRZ}$ (the calibration value is $(12.01 \pm 1.1) \times 10^9$ photons/sr/pC in the normal direction[30]) and the efficiency of the camera $f_{EMCCD}$, which is given by

$$H = f_{DRZ} \cdot f_{optical} \cdot f_{EMCCD} \cdot N_e$$

The $f_{optical}$ is ~0.0048, determined by the collection angle of imaging lens and iris. The $f_{EMCCD}$ is estimated by $f_{EMCCD} = \frac{QE}{AD} \cdot EM_{gain}$ where QE is the quantum efficiency of EMCCD, AD is the CCD sensitivity (electrons per A/D count) under specific pre-amplification rate, and $EM_{gain}$ is the gain of EMCCD. In the experiment, $f_{EMCCD}$ is ~6.79 with $EM_{gain} = 100$.

**Error analysis of lattice constant measurement.**

According to the Bragg diffraction law $dsin\theta = \lambda$, the lattice constant d can be known after the diffraction angle $\theta$ and the de Broglie wavelength of electron probe $\lambda$ are known. For the small-angle approximation, $\theta \approx \tan\theta = R/L$ where $L$ is the imaging distance from sample to screen and $R$ is the offset of the diffraction spots from the direct incident beam without diffraction. The final measurement error of lattice constant $\delta d$ couples the errors from $R$, $L$ and $\lambda$, which can be expressed as

$$\frac{\delta d}{d} = \sqrt{(\delta R/R)^2 + (\delta L/L)^2 + (\delta\lambda/\lambda)^2}$$

according to the error transfer theory. Given the term $\delta L/L$ is negligible and $\delta R/R$ is below 1% (due to the high-resolution optical imaging system), the dominant error originates from energy spectrum measurement. The error term $\delta\lambda/\lambda$ ($\approx \delta E/E$) includes the system error and the statistical error. By carefully calibrating the energy spectrometer considering the magnet geometry and magnetic field measurement, the system error is estimated to be ~1.1%. The spectrum measurement of consecutive 60 shots lead to a statistical error of 0.2%. Therefore, the uncertainty of the de Broglie wavelength of electron probe $\delta\lambda/\lambda$ is ~1.12%. Considering all the above error sources the final measurement error of $d$ is ~1.5%.